\documentclass[aps,prb,twocolumn,10pt]{revtex4-1}

\usepackage{graphicx, bm, amsmath, amsfonts,amssymb,amsthm}

\usepackage{float}
\usepackage{caption}
\usepackage{subcaption}
\usepackage{color}
\usepackage{ulem}

\newcommand{\Z}{\mathbb{Z}}

\newcommand{\la}{\langle}
\newcommand{\ra}{\rangle}

\graphicspath{{figures/}}

\begin{document}

\title{Realizing anomalous anyonic symmetries at the surfaces of 3d gauge theories}
\author{Lukasz Fidkowski}
\affiliation{Department of Physics and Astronomy, Stony Brook University, Stony Brook, NY 11794-3800, USA}
\author{Ashvin Vishwanath}
\affiliation{Department of Physics, University of California, Berkeley, CA, 94720, USA}

\begin{abstract}
The hallmark of a 2 dimensional topologically ordered phase is the existence of deconfined `anyon' excitations that have exotic braiding and exchange statistics, different from those of ordinary bosons or fermions.  As opposed to conventional Landau-Ginzburg-Wilson phases, which are classified on the basis of the spontaneous breaking of an underlying symmetry, topologically ordered phases, such as those occurring in the fractional quantum Hall effect, are absolutely stable, not requiring any such symmetry.  Recently, though, it has been realized that symmetries, which may still be present in such systems, can give rise to a host of new, distinct, many-body phases, all of which share the same underlying topological order.  These `symmetry enriched' topological (SET) phases are distinguished not on the basis of anyon braiding statistics alone, but also by the symmetry properties of the anyons, such as their fractional charges, or the way that different anyons are permuted by the symmetry.  Thus, a useful approach to classifying SETs is to determine all possible such symmetry actions on the anyons that are algebraically consistent with the anyons' statistics.  Remarkably, however, there exist symmetry actions that, despite being algebraically consistent, cannot be realized in any physical system, and hence do not lead to valid 2d SETs.  One class of such `anomalous' SETs, characterized by certain disallowed symmetry fractionalization patters, finds a physical interpretation as an allowed surface state of certain 3d short-range entangled phases, but another, characterized by some seemingly valid but anomalous permutation actions of the symmetry on the anyons, has so far eluded a physical interpretation.  In this work, we find a physical realization for these anomalously permuting SETs as surface theories of certain 3d long-range entangled phases, completing our understanding of general anomalous SETs in 2 dimensions.

\end{abstract}

\maketitle


\section{Introduction}

Recently it has been realized that gapped phases of quantum many body systems can be distinguished on the basis of global symmetries, even when these symmetries are unbroken.  Examples include quantum spin Hall states, 3d topological insulators, as well as a plethora of theoretically predicted interacting `symmetry protected topological' (SPT) states\cite{Chen1d, Fidkowski1d, TurnerBerg, Chen2d, Chen2013, DW, LevinGu, WangLevin}.  Despite the lack of an order parameter, all of these form novel symmetry-protected many body quantum phases, and thus transcend the Landau-Ginzburg-Wilson symmetry breaking paradigm.  Similarly, unbroken symmetries can also in principle distinguish among more exotic gapped phases, namely those of topologically ordered systems.  For example, there exist, in principle, multiple distinct phases for the Kagome chiral spin liquid\cite{ZV} that are distinguished only by the different patterns of space group symmetry fractionalization on the anyons.  This makes the classification of such `symmetry-enriched' topological phases\cite{Essin2013, Mesaros2013} an interesting problem.

One approach to understanding two dimensional bosonic symmetry-enriched topological (SET) phases is to gauge the symmetry, which is assumed to be finite, onsite, and unitary, and examine the resulting topological order.  The problem of classifying SETs then turns into the problem of classifying the various topological orders that can result from gauging, and this turns out to be tractable\cite{Maissam2014, Teo_Hughes, Tarantino2015, ENO}.  Very roughly, the result of this analysis is that there are three ways in which two SETs with the same anyons can differ: 1) the symmetry can permute the anyons in the SET - as happens, for example, in a bilayer FQH system under the $\Z_2$ symmetry that exchanges layers - and two SETs can differ in the permutation action of the symmetry, 2) even if there is no permutation action, the anyons can carry fractional symmetry charges - as happens, e.g. in ordinary Laughlin states, or in the chiral spin liquid - and two SETs can differ in the assignment of such fractional charges, and 3) even if there are no permutations or fractional charges, two SETs can differ in that one can be obtained from the other by the stacking of an SPT.

Interestingly, although every SET corresponds to some choice of the three pieces of data above, not every such choice leads to a valid SET.  For a concrete example, consider the topological order of the chiral spin liquid, which has only one non-trivial anyon, the spinon, and break the $SO(3)$ spin rotation symmetry down to the $\Z_2 \times \Z_2$ subgroup consisting of $180$ degree rotations around the principal axes.  For this discrete group, it turns out that, in addition to the usual Kalmeyer-Laughlin spin liquid, there exist three other seemingly valid patterns of $\Z_2 \times \Z_2$ symmetry fractionalization on the spinon.  However, these seemingly allowed fractionalization patterns are actually anomalous, in that there is no SET that realizes them.

Despite the lack of a 2d SET realization, the algebraic consistency of these anomalous fractionalization patterns suggests that they should have some physical interpretation.  Indeed, in \onlinecite{ProjS} it was shown that the anomalous chiral spin liquids, although not realizable strictly in 2d, can be realized as surface states of 3d SPTs.  These novel symmetric gapped surface terminations broadened the class of known SPT surface states, and spurred the discovery of similar topologically ordered symmetric surface terminations for the ordinary topological insulator\cite{Chen2013a}, as well as 3d topological superconductors\cite{Fidkowski2013}.  Furthermore, beyond the particular chiral spin liquid example, they led to a general conjecture relating anomalous symmetry fractionalization in general anyon theories and 3d SPT realizations.

Beyond the fractionalization anomaly, however, there is one other case in which the SET data defined above fails to lead to a valid SET \cite{ENO}, which has so far resisted a physical interpretation.  Namely, certain seemingly valid permutations of the anyons by the symmetry action (`anyonic symmetries' in the language of reference \onlinecite{Teo_Hughes}) are impossible to realize in 2d SETs.  Again, because these permutations are compatible with the fusion and braiding structure of the anyons, one expects that, despite the lack of any 2d SET realization, they should have some physical interpretation.

In this paper we find precisely such a physical interpretation.  Namely, we show that at least a certain subclass of such obstructed symmetry actions can be realized at the surface of 3d SETs.  These 3d SETs are special, because they exhibit a novel type of symmetry fractionalization along their loop-like excitations rather than point-like particles, a new type of symmetry fractionalization available only in 3 dimensions \cite{HermeleChen}.  It is interesting that, just as the previously discussed fractionalization anomaly for 2d SETs was related to surface realizations on intrinsically 3d SPT phases, so the permutation anomaly is related to surface realizations on intrinsically 3d SET phases.

The rest of the paper is structured as follows.  In section \ref{overview} we give an overview of our construction for one concrete example of the permutation anomaly.  In section \ref{sec:surface} we describe the general class of permutation-anomalous surface theories in more generality.  In section \ref{sec:bulk} we discuss the bulk 3d SET which realizes our anomalous surface theory, and construct an exactly solvable model for it.  In section \ref{sec:combined}, we combine the bulk model constructed in section \ref{sec:bulk} with the anomalous surfaces discussed in section \ref{sec:surface}, and in section \ref{sec:gen} we discuss some generalizations of our constructions and present a conjecture.  Finally, we conclude in section \ref{sec:conc} with some ideas about how to extend this work, in particular to fermionic SPTs and topological orders.

\section{Concrete example of the permutation anomaly}\label{overview}
\begin{figure*}[htbp]
\begin{center}
\includegraphics[width=0.6\textwidth]{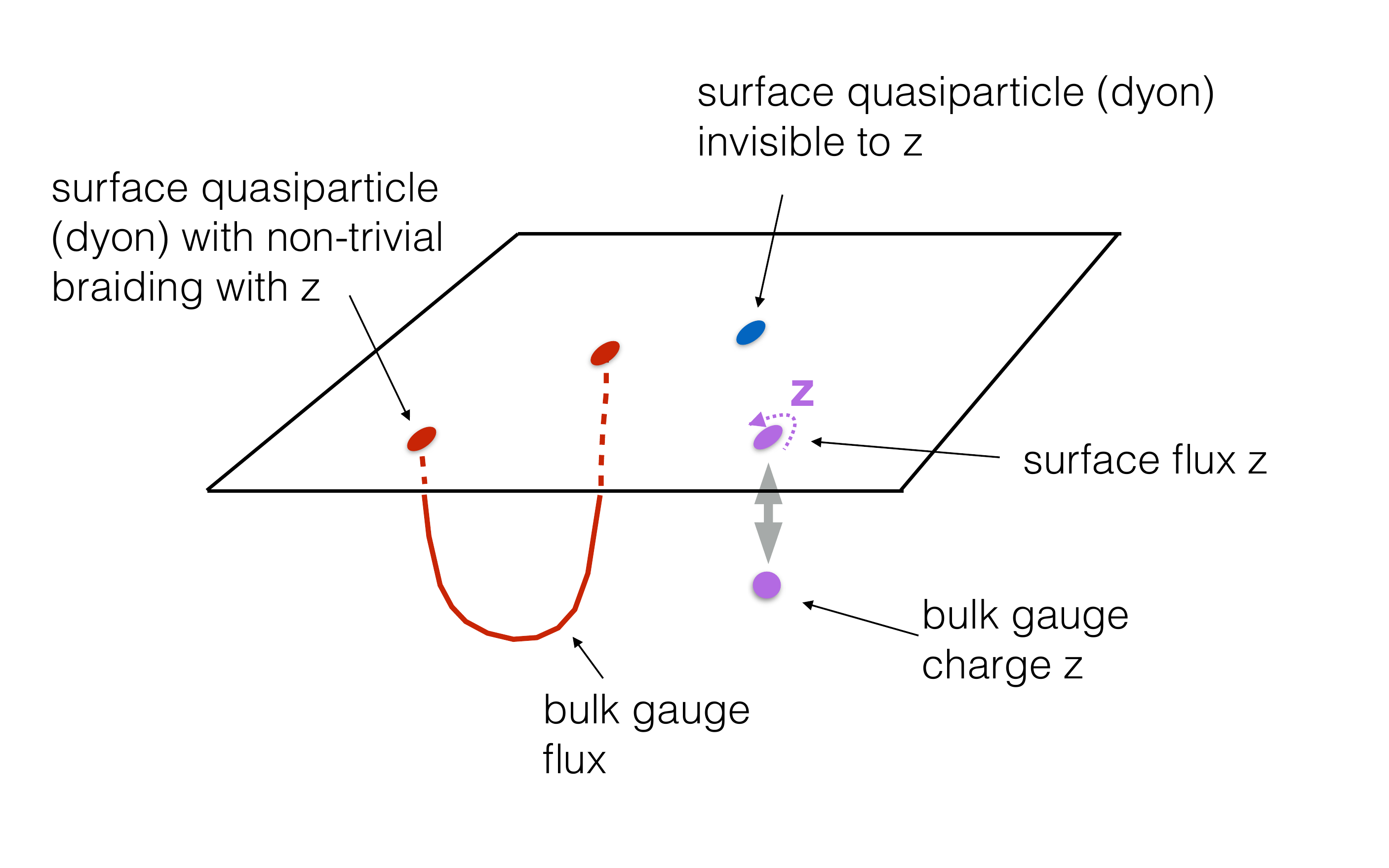}
\caption{Excitations in the bulk and surface of our model.  The bulk contains $\Z_2$ gauge charges (purple) and gauge flux loops (red).  When the bulk gauge charges come up to the surface, they are identified with fluxes of the $\Z_2$ center of ${\mathbb{D}}_{16}$.  Surface quasiparticles which have trivial full braiding with such a center flux - namely dyons where this center acts trivially in the charge part - are deconfined on the surface (blue), while the other surface quasiparticles (red) are confined to the endpoints of the bulk gague flux loops.  This is because the braiding of the surface flux z around such a quasiparticle must reproduce the same phase of $-1$ as the braiding of the bulk gauge charge around the bulk gauge charge loop.  In the general setting, the surface is a gauge theory of a non-abelian group with center $Z$, and the bulk is a $Z^*$ gauge theory.}
\label{fig1}
\end{center}
\end{figure*}

A concrete example, on which we focus in this paper, is the following.  For the anomalous 2d surface, we take a 2d discrete gauge theory (or `quantum double') of the finite non-abelian group ${\mathbb D}_{16}$, which just consists of the spatial symmetries of the octagon, generated by a reflection and 45 degree rotation.  For the symmetry group, we take $G=\Z_2^G$.  To define the action of $G$ on the quasiparticles of this gauge theory, we use some purely mathematical facts about ${\mathbb D}_{16}$.  Namely, ${\mathbb D}_{16}$ has an outer automorphism ${\hat{\sigma}}$ - i.e. a relabeling of the group elements, respecting the group law, which is not given by conjugating by any element of ${\mathbb D}_{16}$ - whose square ${\hat{\sigma}}^2$ is an inner automorphism (i.e. is given by conjugation by some element $x$).  Promoted to the level of the gauge theory, ${\hat{\sigma}}$ gives a $\Z_2^G$ action on the quasiparticle spectrum, because applying it twice yields the gauge transformation by $x$.  The important fact about ${\hat{\sigma}}$, that results in this $\Z_2^G$ action being anomalous, is that there is no way to extend ${\mathbb D}_{16}$ to a larger group $E$, twice the size of ${\mathbb D}_{16}$, which contains an element $\sigma$ such that conjugating by $\sigma$ is the same as applying ${\hat{\sigma}}$.  Note that this is not a general property of outer automorphisms - there is something very special about our ${\hat{\sigma}}$, which we elucidate below.  The lack of a suitable group extension $E$ can physically be thought of as an impossibility of gauging $G$, if we assume the result to be another discrete gauge theory; a more formal argument shows that with this action of $G$, the ${\mathbb D}_{16}$ gauge theory does indeed suffer from the $H^3(G,{\cal{A}})$ anomaly defined in references \onlinecite{ENO, Maissam2014} (here ${\cal{A}}$ is the subgroup of abelian surface anyons).

For the bulk 3d SET, we take a 3d $\Z_2$ gauge theory (i.e. 3d toric code).  Under the $\Z_2^G$ action, the $\Z_2$ gauge charges are taken to transform trivially, but the $\Z_2$ gauge fluxes - or `vison' loops - are taken to transform like the edge of a non-trivial $\Z_2^G$ SPT.  There are various ways of constructing such a 3d SET: for example one could gauge $\Z_2$ in the decorated domain wall \cite{Chen2013b} or Walker-Wang \cite{ProjS} model of a 3d $\Z_2 \times \Z_2^G$ SPT.  However, we choose a different route, and instead, motivated by a model of Hermele \cite{Hermele_string_flux}, construct an exactly solved Hamiltonian model of membranes fluctuating in 3 dimensions.  This membrane representation is dual to the usual electric string representation of the 3d toric code, and allows us to directly access the symmetry fractionalization on the vison loops (which here are the edges of the membranes).  We will also be able to realize the anomalous ${\mathbb D}_{16}$ gauge theory as an exactly solvable surface termination of this model, explicitly illustrating our claim above.  A feature of this construction is that when the bulk $\Z_2$ gauge charge comes up to the surface, it is identified with the surface flux of the $\Z_2$ center of ${\mathbb D}_{16}$, which forces certain surface ${\mathbb D}_{16}$ charges - i.e. those which transform non-trivially under the center - to be confined to the endpoints of vison loops.

Going beyond our specific construction, it is instructive to examine other surface terminations of this 3d SET.  Broadly speaking, there are two classes: one can either (I) condense the $\Z_2$ gauge charges at the surface, or (II) condense the visons.  As we show, the former can actually be gapped and symmetric without any additional surface topological order.  One way to understand this is to notice that vison loops cannot end at this surface, since vison endpoints are confined due to the surface Higgs condensate.  Since these visons loops are the only objects carrying non-trivial fractionalization under $\Z_2^G$ it is natural to expect that a trivial gapped surface should be allowed in this case.  On the other hand, our ${\mathbb D}_{16}$ termination does allow surface deconfined vison endpoints, and falls in class (II).  One might be tempted to conjecture that, just like our ${\mathbb D}_{16}$ surface, all gapped symmetric surfaces in class (II) suffer from the $H^3(G,A)$ anomaly, but this is not quite correct.  Indeed, a counterexample comes from gauging the $\Z_2$ in the 3d $\Z_2 \times \Z_2^G$ parent SPT with the projective semion surface state \cite{Bi2013, ProjS}.  The result is our 3d SET with a $U(1)_8$ surface state, with the odd truncated $U(1)$ charges bound to the vison endpoints.  Here the $\Z_2^G$ symmetry just reverses the truncated $U(1)$ charge: $m \rightarrow -m$, where $m$ is an integer modulo $8$.  However, such a symmetry action can certainly occur in a purely 2d realization of $U(1)_8$, as can be seen e.g. by gauging the $\Z_2$ in the 2d $\Z_2 \times \Z_2^G$ chiral spin liquid \cite{ProjS, Teo_Hughes}.

How can the same 3d SET support both an anomalous and a seemingly non-anomalous surface state?  The mathematical resolution to this puzzle is that the bulk SET order and surface anomaly cannot be matched up directly because they take values in different groups: the former in $H^3(\Z_2^G, \Z_2)$, and the latter in $H^3(\Z_2^G,\Z_8)$ for the $U(1)_8$ surface.  In order to compare them, we have to use the fact that the bulk $\Z_2$ gauge charge is identified with the $m=4$ boson of the surface theory to embed $\Z_2$ as a subgroup of $\Z_8$.  However, under this embedding, the non-trivial cohomology class in $H^3(\Z_2^G, \Z_2)$ becomes zero in $H^3(G,\Z_8)$.  Said another way, the 3-cocycle that describes this cohomology class can be gauged away with $\Z_8$ coefficients, but not with $\Z_2$ coefficients.  Physically, this is because the $U(1)_8$ anyons come in $1/4$ fractions of the $m=4$ gauge charge, and binding one such $m=1$ anyon to a $\Z_2^G$ symmetry flux allows one to `screen' the anomaly, as we will see below \footnote{More formally, this screening corresponds to a 2-cochain of $\Z_2^G$, and involves a modification of the $\Z_2^G$ flux fusion rule by $m=2$}.  Thus, physically, the resolution to our puzzle is that both the ${\mathbb D}_{16}$ and the $U(1)_8$ surface states are anomalous, but in the latter the anomaly is only seen at the level of flux fusion rules, and not the $H^3(\Z_2^G,{\cal{A}})$ cohomology group.

The ${\mathbb D}_{16}$ example worked out in this paper can be generalized.  Indeed, it is one example of the so-called Eilenberg-Maclane obstruction in group theory, where ${\mathbb D}_{16}$ and its center $\Z_2$ are replaced by an arbitrary non-abelian finite group $H$ with a center $Z$.  The symmetry group $G$ then acts by automorphisms of $H$, but the group relations are only required to close modulo conjugation by group elements of $H$.  What is obstructed is again the existence of a suitable group extension $E$, and once again this has a natural physical interpretation: $G$, promoted to a permutation symmetry of the $H$ gauge theory quasiparticle spectrum, suffers from the $H^3(G,A)$ obstruction and cannot be realized in any gapped 2d system \cite{ENO}.  The exactly solved bulk and surface models written down in this paper can be generalized to this setting; note that the 3d bulk is still just a gauge theory of a finite abelian group, namely $Z$.

Motivated by these examples, it is tempting to conjecture a general relation between the 2d $H^3(G,{\cal{A}})$ obstruction and 3d SETs.  However, an immediate problem arises: in general, we do not know if the 3-cocycle representing the obstruction class can be taken to be valued in some subset of abelian anyons that happen to all be mutual bosons.  If it cannot, then these abelian anyons cannot be identified with the quasiparticles of any 3d SET.  Note that in principle some of the quasi-particles could be self-fermions.  In particular, one might have a situation with a single fermion in the 3d bulk, which would correspond to a gauged version of a 3d fermionic SPT.  If the symmetry group $G$ could be generalized to include anti-unitary elements then one example of this might be the $SO(3)_3$ surface topological order proposed for odd $\nu$ topological superconductors in Cartan class AIII.  However, more work needs to be done to make any of these ideas concrete.

\section{Anomalous anyonic symmetries} \label{sec:surface}

We begin by considering symmetries of 2d discrete gauge theories.  While ultimately it will be crucial for us to consider non-abelian gauge groups $H$, let us first review the simpler case of an abelian gauge group, which for clarity we denote $A$.  The spectrum of topological excitations of such a 2d gauge theory with finite abelian gauge group $A$ contains charges, fluxes, and charge-flux composites.  If we denote by $(q,a)$ a charge-flux composite with charge $q$ and flux $a$, then the exchange statistics of an excitation $(q,a)$ are given by the phase $\la a, q\ra$ while the mutual statistics of $(q,a)$ and $(q',a')$ are given by $\la a, q'\ra + \la a', q\ra$.  Here $\la a, q \ra$ refers to the phase, modulo $2\pi$, obtained by acting with the group element $a$ on the irreducible representation $q$.  For example, if $A=\Z_n$, then both $a$ and $q$ can be thought of as integers modulo $n$, and $\la a, q \ra = 2\pi aq/n$.

Now consider how a finite symmetry group $G$ can act on this spectrum of excitations.  One such set of symmetry actions can be obtained from permutations $\rho_g$ of $A$ which preserve the group law, which we will refer to as {\emph{automorphisms}} of $A$.  Indeed, an automorphism $a \rightarrow \rho_g \cdot a$ gives an action on both the fluxes, which are just group elements of $A$, as well as the charges, which are irreducible representations of $A$, and it is easy to see that the statistics of these excitations are unchanged under such an action.  For example, when $A=\Z_n$ all automorphisms are of the form $a \rightarrow ma$ where $m$ is an integer relatively prime to $n$, and the action of such an automorphism on a charge-flux composite $(q,a)$ is $(q,a)\rightarrow (m^{-1} q,ma)$, where by $m^{-1}$ we mean an {\emph{integer}} satisfying $m m^{-1} = 1$ modulo $n$.  The action of any automorphism $\rho_g$ of any abelian $A$ can actually be realized as a microscopic onsite symmetry in a lattice model of an $A$ gauge theory - see, reference \onlinecite{Tarantino2015} for a specific construction.  Thus there are no obstructions to realizing such symmetries for abelian discrete gauge theories. 

The situation is more complicated for non-abelian gauge groups $H$.  For one thing, there is now a special class of automorphisms of $H$, given by $h\rightarrow xhx^{-1}$ for some fixed $x\in H$.  These are called `inner' automorphisms, and they are special because they do not permute the quasiparticles of the $H$ gauge theory \footnote{More precisely, they induce a trivial braided tensor automorphism of the quasiparticles.}.  This is because such an inner automorphism is nothing more than a global gauge transformation by $x\in H$; for example, recall that the pure flux quasiparticles in a non-abelian $H$ gauge theory correspond to conjugacy classes of $H$, which are by definition invariant under $h\rightarrow xhx^{-1}$.  Thus, when one defines a group action of $G$ in the non-abelian gauge group setting, it is appropriate to consider automorphisms $\rho_g$ which obey the composition law in $G$ only up to inner automorphisms:

\begin{equation} \label{aut_cond}
\rho_{g_1} \cdot \rho_{g_2} \cdot h = x_{g_1,g_2} (\rho_{g_1 g_2} \cdot h) x_{g_1,g_2}^{-1}.
\end{equation}
for some $x_{g_1,g_2}\in H$.  This is because the conjugation by $x_{g_1,g_2}$ does not do anything to the $H$ gauge theory quasiparticles, so that the permutation actions induced by the $\rho_g$ on these quasiparticles satisfy the $G$ group law exactly.  Just as in the abelian case, it can be checked that this $G$ action preserves the statistics of these quasiparticle excitations.

\begin{figure}[htbp]
\includegraphics[width=0.45\textwidth]{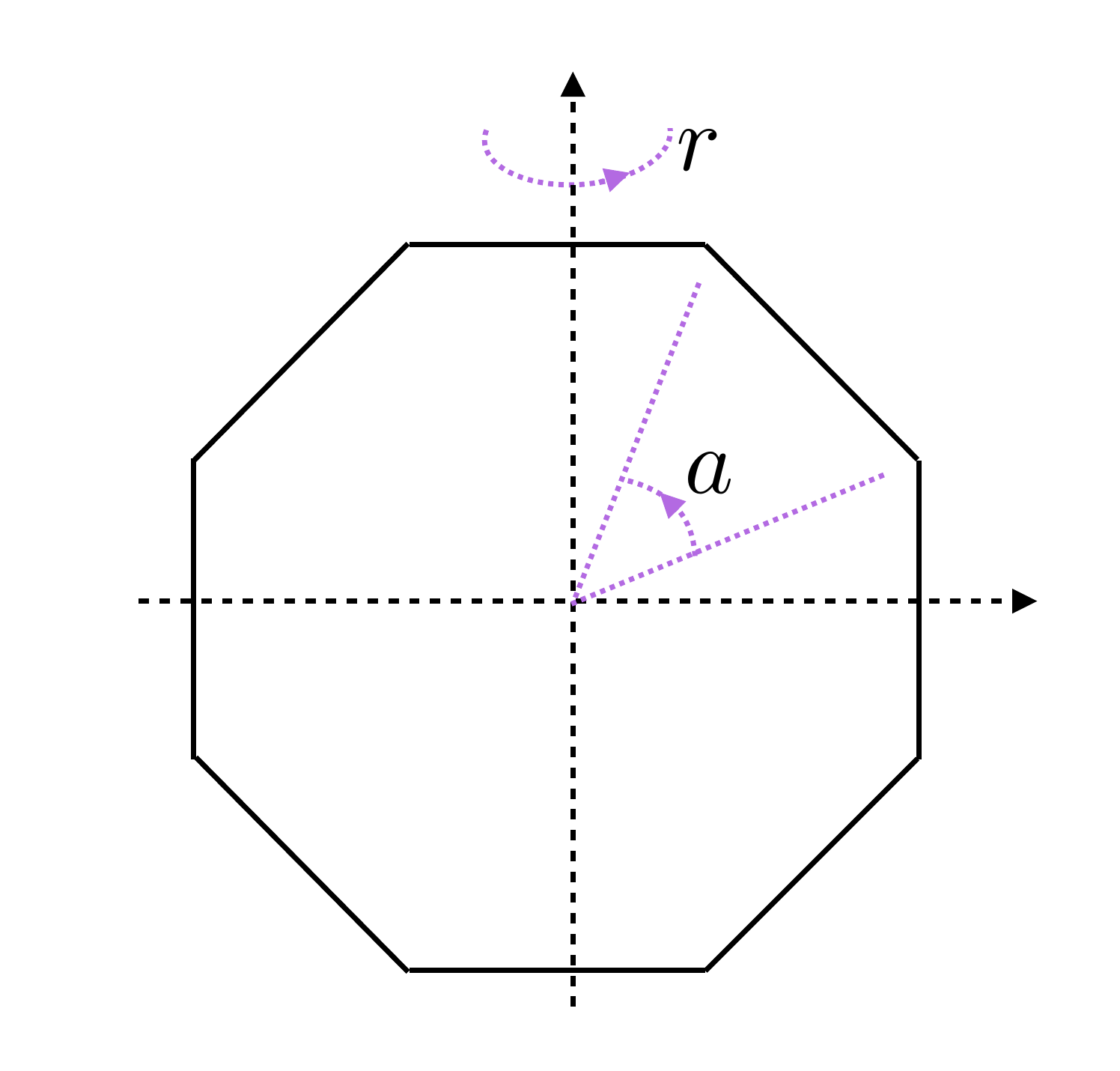}
\caption{The generators $r$ and $a$ of ${\mathbb{D}_{16}}$, the group of spatial symmetries of the regular octagon.}
\label{fig4}
\end{figure}

One might thus expect that, because all of the statistics are preserved, such an action of $G$ can be realized in some microscopic 2d model of the $H$ gauge theory, with $G$ acting as an onsite symmetry.  However, this turns out not to be the case: there exists an invariant that is a function of the data in the problem - namely $H$, $G$, and $\{\rho_g\}$ - such that, when this invariant is non-trivial, a microscopic symmetric 2d realization is impossible.  Specifically, the invariant is a function $z(g_1,g_2,g_3)$ of three group variables $g_1,g_2,g_3\in G$, valued in $Z$, the center of $H$.\footnote{Recall that the center of a group $H$ is defined as the set of all $z\in H$ which commute with all of $H$, i.e. $zh=hz$ for all $h\in H$.}  This function turns out to always satisfy the so-called co-cycle condition:

\begin{align} \label{cocyc_z}
z(g_2,g_3,g_4)&-z(g_1 g_2,g_3,g_4)+z(g_1,g_2 g_3,g_4)\\
  &-z(g_1,g_2,g_3 g_4)+z(g_1,g_2,g_3)=0.
\end{align}
Furthermore, there is a notion of gauge transformation: $z$ and $z'$ are considered gauge equivalent if:

\begin{align} \label{gauge_z}
z'(g_1,g_2,g_3)&=z(g_1,g_2,g_3)+\eta(g_2,g_3) -\eta(g_1 g_2,g_3) \\ &+\eta(g_1,g_2 g_3)-\eta(g_1,g_2)
\end{align}
for some $Z$ valued function $\eta$ of two group variables.  As we show in appendix \ref{group_obstruction}, it is precisely when $z(g_1,g_2,g_3)$ is not gauge equivalent to $0$, in the sense of eq. \ref{gauge_z}, that a microscopic symmetric 2d realization is impossible and the anyonic symmetry is anomalous.  The set of gauge equivalence classes so defined is called the third cohomology group of $G$, and is denoted $H^3(G,Z)$; the anomaly is thus signaled by a non-zero class $[z] \in H^3(G,Z)$.

The rough idea behind the argument that anyonic symmetries with non-zero $[z]$ are not realizable strictly in 2d is to suppose for a contradiction that they were, and gauge $G$.  Then we might expect that the result would be a gauge theory of a larger group $E$ that contains $H$ as a normal subgroup, and such that $E/H=G$: this is a generalization of a twisted product of $G$ and $H$.  Now, for each $g\in G$, pick a lift to $x_g \in E$.  Then $E$ must have the property that conjugation by $x_g$ induces the automorphism $\rho_g$ of $H$, up to an inner automorphism of $H$.  However, it turns out that mathematically this is not always possible, and the obstruction is parametrized by $[z] \in H^3(G,Z)$ - see appendix \ref{group_obstruction} for details.

This argument can be made even more concrete for the case of $H={\mathbb{D}}_{16}$.  As discussed below, ${\mathbb{D}}_{16}$ is the group of symmetries of the regular octagon, generated by a $45$ degree rotation $a$ and a reflection $r$ - see figure \ref{fig4}.  The group relations are $rar = a^{-1}$, and $a^8=1$.  The symmetry group $G$ is taken to be $\Z_2^G = \{1,\sigma\}$, with $G$ acting by $r\rightarrow ra$ and $a\rightarrow a^5$.  This generates a valid automorphism of ${\mathbb{D}}_{16}$.  As discussed below, this automorphism squares to an inner automorphism of $H$, namely conjugation by $a^3$, so we have a valid group map from $G$ to the group of automorphisms of $H$ modulo inner automorphisms.  But there is no extension $E$ that can exist in this case.  Indeed, suppose for a contradiction that such an $E$ did exist, and let $s$ denote a lift of $\sigma$ to $E$.  Then, since $E$ is the disjoint union of the two cosets $H$, $sH$, we must have $s^2\in H$, and, furthermore, conjugation by $s^2$ must be the same as conjugation by $a^3$ (on $H$).  This means that either $s^2=a^3$ or $s^2=a^{-1}$.  Suppose first that $s^2=a^3$, and consider the triple product $sss$.  On the one hand,

\begin{equation}
sss=s^2 s=a^3 s=s a^{-1}
\end{equation}
On the other hand,
\begin{equation}
sss=s s^2=sa^3
\end{equation}
Since these two are not equal, we have a contradiction.  The same argument applies if $s^2=a^{-1}$.

A more rigorous argument, for the general discrete gauge theory case, is given in the appendix to reference \onlinecite{ENO}.  There, it is shown that the cohomology class $[z] \in H^3(G,Z)$ constructed in appendix \ref{group_obstruction} is the same as the $H^3(G,{\cal{A}})$ obstruction to extending an action of $G$ by braided autoequivalences on an arbitrary modular tensor category defined in reference \onlinecite{Maissam2014}.  Here ${\cal{A}}$ is the set of abelian anyons in the theory, of which the center fluxes $Z$ form a subset.  In fact, ${\cal{A}}$ is of the form $Z \times Z'$, which causes the map $H^3(G,Z) \rightarrow H^3(G,{\cal{A}})$ induced by the inclusion to be one to one.

Next we want to write down a symmetric lattice 2d Hamiltonian for the ${\mathbb{D}}_{16}$ gauge theory.  Since this theory is anomalous, we cannot write down a fully gapped such Hamiltonian; rather, the Hamiltonian we write down will have extensive degeneracy.  Later, we will remove this degeneracy by coupling to the bulk 3d SET.  As a warmup, we first write down a Hamiltonian for a related non-anomalous theory. 

\subsection{Model of a 2d discrete non-abelian gauge theory with non-anomalous anyonic symmetry}

Our warm up non-anomalous theory will be the discrete gauge theory of ${\mathbb{D}}_8$, which is a group of order $8$ that can be thought of as the group of spatial symmetries of the square.   We will denote the generators of this group ${\tilde{a}}, r$; they satisfy the relations ${\tilde{a}}^4=r^2=1$, $r{\tilde{a}}r={\tilde{a}}^{-1}$.  The $\Z_2^G$ symmetry group acts here by the automorphism $\tilde{\rho}$, defined by $r\rightarrow r{\tilde{a}}, {\tilde{a}}\rightarrow{\tilde{a}}$.  It can be checked that this symmetry is not anomalous.  Indeed, the appropriate extension of ${\mathbb{D}}_8$ by $\Z_2^G$ is just given by ${\mathbb{D}}_{16}$, with ${\mathbb{D}}_8$ embedded as $r=r, {\tilde{a}}=a^2$.  This is because conjugation by $a^{-1}$ gives precisely the group action defined above.  Thus, in this case we expect  a purely 2d realization, and this is what we write down in this subsection in the form of an exactly solved model.

\begin{figure*}[htbp]
\begin{center}
\includegraphics[width=0.55\textwidth]{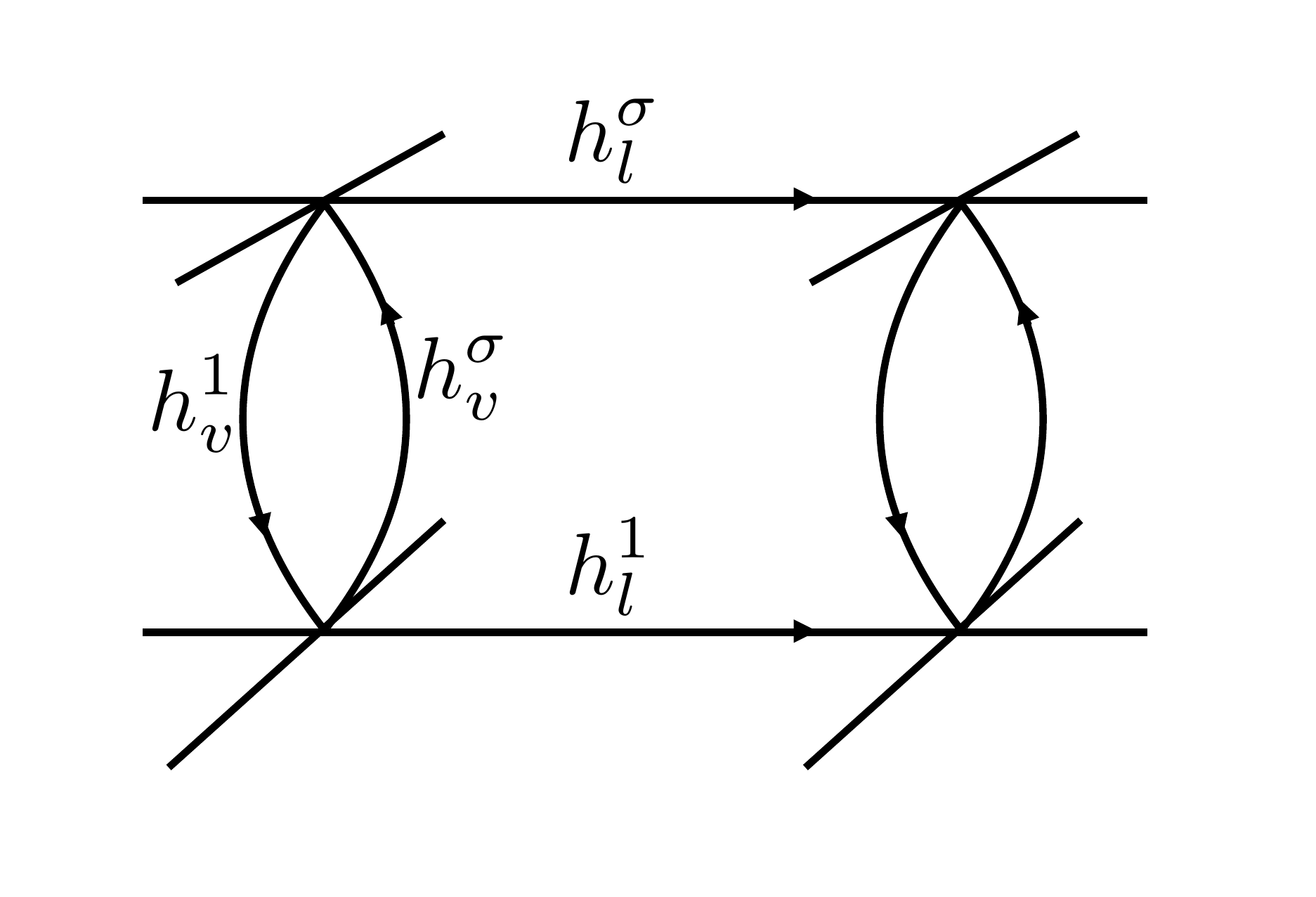}
\caption{Quasi 2d lattice $\Gamma^s$ on which the ${\mathbb{D}}_8$ gauge theory is defined.  Here $h^{1,\sigma}_v$ and $h^{1,\sigma}_l$ are ${\mathbb{D}}_8$ variables associated with `horizontal' and `vertical' links respectively.  The same lattice will be used to define the anomalous ${\mathbb{D}}_{16}$ gauge theory, though we we will need to introduce additional bulk degrees of freedom to realize a fully gapped and symmetric model.}
\label{fig3}
\end{center}
\end{figure*}

We start by defining a certain oriented quasi-2d lattice $\Gamma^s$ - see figure \ref{fig3} - on whose links we will have ${\mathbb{D}}_8$ group labels as our degrees of freedom. $\Gamma^s$ consists of two copies (the $1$ and $\sigma$ copies) of the square lattice, with nearest neighbor links $l^1, l^{\sigma}$ oriented in the positive $x$ and $y$ directions, together with additional links $l_v^1$ and $l_v^{\sigma}$ connecting vertices in the two copies, where $v^g$ is a vertex in copy $g$.  This is precisely the lattice used by reference \onlinecite{Hermele_string_flux} to construct different fractionalization patterns for a $\Z_2^G$ enriched toric code.  Our construction is in some sense a non-abelian extension of that of references \onlinecite{Hermele_string_flux, Tarantino2015}.

We take an $H={\mathbb{D}}_8$ gauge theory defined on this quasi-2d lattice $\Gamma^s$.  That is, each `horizontal' link $l^{g}$ ($g=1,\sigma$) carries an $H$ label $h_l^g$, and each `vertical' link $l_v^g$ carries an $H$ label $h_v^g$ - see figure \ref{fig3}.  The Hamiltonian contains vertex terms $A''_{v,g}$ which enforce Gauss's law at each vertex $v_g$, together with plaquette terms $B''_{v,g}, B''_{l,g}, B''_{p,g}$ which enforce zero flux through various types of plaquettes in the lattice:

\begin{align} \label{ham_2d}
H_{{\mathbb{D}_8}}=-\sum_{v,g} A''_{v,g} - \sum_{v,g} B''_{v,g} - \sum_{l,g} B''_{l,g} - \sum_{p,g} B''_{p,g}.
\end{align}

Here
\begin{align}
A''_{v,g}=\sum_{h\in H} {A''}^{(h)}_{v,g}
\end{align}
and ${A''}^{(h)}_{v,g}$ is defined as the gauge transformation by $h$ at vertex $v^g$.  More precisely, ${A''}^{(h)}_{v,g}$ acts on a gauge field configuration $\{ h_l^g, h_v^g\}$ by

\begin{align}
{A''}^{(h)}_{v,g}: & h_l^{g} \rightarrow h h_l^g \,\,\,\,\text{ for } v \rightarrow l \\
& h_l^{g} \rightarrow h_l^g h^{-1}  \,\,\,\,\text{ for } v \leftarrow l \\
& h_v^{g} \rightarrow h h_v^g \\
& h_v^{\sigma g} \rightarrow h h_v^{\sigma g}
\end{align}
where the notation $v \rightarrow l$ means $l$ is oriented away from $v$, and $v \leftarrow l$ means $l$ is oriented towards $v$.

The precise definitions of the plaquette terms are:

\begin{align}
{B''}_{v,g} = \,\, &1 \,\,\,\, \text{ if } h_v^{1} h_v^{\sigma} = 1 \\
& 0 \,\,\,\, \text{ otherwise},
\end{align}

\begin{align}
{B''}_{l,g} = \,\, &1 \,\,\,\, \text{ if } h_v^{g} h_l^{\sigma g} (h_{v'}^g)^{-1} (h_l^{g})^{-1} = 1 \\
& 0 \,\,\,\, \text{ otherwise},
\end{align}
where $l=\la v v' \ra$, and

\begin{align}
{B''}_{p,g} = \,\, &1 \,\,\,\, \text{ if } \prod_{l\in p} (h_l^g)^{s_p(l)} = 1 \\
& 0 \,\,\,\, \text{ otherwise},
\end{align}
where $s_p(l)=\pm 1$ depending on whether $l$ is oriented with or against the counterclockwise orientation of the plaquette $p$.  These are just the usual plaquette terms which energetically prefer zero flux through each plaquette.

So far we have just constructed a standard model of ${\mathbb{D}}_8$ gauge theory on the quasi 2d lattice $\Gamma^s$.  Now we will define the action of our $\Z_2^G$ as follows.  The non-trivial generator $\sigma$ of $G$ will first of all act on $\Gamma^s$ by exchanging the links $l^g \leftrightarrow l^{\sigma g}$ and $v^g \leftrightarrow v^{\sigma g}$.  However, it does not simply exchange the ${\mathbb{D}}_8$ labels on these links.  Rather, we define:

\begin{align} \label{Gaction}
h_v^{1} & \rightarrow {\tilde{a}} (\tilde{\rho} \cdot h_v^{\sigma}) \\
h_v^{\sigma} & \rightarrow (\tilde{\rho} \cdot h_v^1) {\tilde{a}}^{-1} \\
h_l^1 & \rightarrow {\tilde{a}} (\tilde{\rho} \cdot h_v^{\sigma}) {\tilde{a}}^{-1} \\
h_l^{\sigma} & \rightarrow \tilde{\rho} \cdot h_v^{1}.
\end{align}
Recall that the automorphism $\tilde{\rho}$ acts by $r \rightarrow r {\tilde{a}}$, $\tilde{a} \rightarrow \tilde{a}$.  Using the fact that applying the symmetry on the ${\mathbb{D}}_8$ labels twice is the same as conjugation by ${\tilde{a}}^{-1}$, we see that applying the $G$ action defined in eq. \ref{Gaction} twice yields the identity, so that this defines a valid onsite microscopic $\Z_2$ symmetry (where the sites are actually supersites that include two links each).  Furthermore, it is straightforward to verify that this action commutes with all of the terms in the Hamiltonian.  Indeed, the action defined in eq. \ref{Gaction} is simply given by applying the symmetry and then performing a gauge transformation by $a\in H$ on copy $1$ after doing the exchange: this clearly commutes with all of the vertex and plaquette terms in the Hamiltonian in eq. \ref{ham_2d}.  Also for this reason, the ${\mathbb{D}}_8$ fluxes just transform according to the symmetry insofar as the effective $2d$ ${\mathbb{D}}_8$ gauge theory is concerned - the gauge transformation does not do anything.  We have thus built a quasi 2d model which realizes the ${\mathbb{D}}_8$ topological order, in which the $\Z_2^G$ symmetry acts onsite.

\subsection{Example of a non-abelian gauge theory with an obstructed symmetry action}

In the previous subsection we considered ${\mathbb{D}}_8$ gauge theory together with a certain $\Z_2^G$ symmetry, and constructed an exactly solved Hamiltonian realizing this $\Z_2^G$ as a microscopic onsite symmetry.  In this section we will give an example of a non-abelian gauge theory with an action of $\Z_2^G$ on its quasiparticles which preserves all of the statistics but nevertheless cannot be realized as microscopic symmetry in 2d - i.e. an anomalous anyonic symmetry.  Here the gauge group will be ${\mathbb{D}}_{16}$, the group of spatial symmetries of an octagon, and the $\Z_2^G$ action will again be defined by a certain automorphism - i.e. permutation preserving the group law - of ${\mathbb{D}}_{16}$ , which we refer to as $\rho$.  Again, the generators of ${\mathbb{D}}_{16}$ are $a$, corresponding to the $45$ degree counterclockwise rotation, and $r$, corresponding to a reflection.  The group relations are $a^8=r^2=1$ and $rar=a^{-1}$.  The automorphism $\rho$ is defined by:

\begin{align}
\rho:&a \rightarrow a^5 \\
 &r \rightarrow ra
 \end{align}
Note that the quotient of ${\mathbb{D}}_{16}$ by its center $\{1,a^4\}$ is precisely ${\mathbb{D}}_8$, and the action of $\rho$ descends to this quotient and is given just by the symmetry action defined in the previous section.  Also note that applying $\rho$ twice gives an inner automorphism of ${\mathbb{D}}_{16}$, namely conjugation by $a^{-3}$.  Thus $\rho$ generates a well defined $\Z_2^G$ action on ${\mathbb{D}}_{16}$ gauge theory.  As we demonstrated above and in appendix \ref{group_obstruction}, this symmetry is anomalous.  However, for now let us just see what fails if we do the naive thing and try to realize this symmetry in the same kind of microscopic model that we used in the previous subsection for $H={\mathbb{D}}_8$, with the exact same lattice $\Gamma^s$, which now carries ${\mathbb{D}}_{16}$ group labels.  

Since $\rho$ squares to conjugation by $a^{-3}$, we can attempt, in analogy with the previous subsection, to define the $\Z_2^G$ action on the gauge theory degrees of freedom by

\begin{align} \label{Gaction}
h_v^{1} & \rightarrow a^3 (\rho \cdot h_v^{\sigma}) \\
h_v^{\sigma} & \rightarrow (\rho \cdot h_v^1) a^{-3} \\
h_l^1 & \rightarrow a^3 (\rho \cdot h_v^{\sigma}) a^{-3} \\
h_l^{\sigma} & \rightarrow \rho \cdot h_v^{1}.
\end{align}
However, there is a problem with this: applying $\rho$ twice to, say, $h_v^1$ we get:

\begin{align}
h_v^{1} \rightarrow \left( \rho \cdot (a^3 (\rho \cdot h_v^{1})) \right) a^{-3}=a^4 h_v^{1}
\end{align}
So this is not a valid microscopic onsite $\Z_2^G$ symmetry: it does not square to the identity.  One can attempt to fix this problem by redefining:

\begin{align} \label{newGaction}
h_v^{1} & \rightarrow a^3 (\rho \cdot h_v^{\sigma}) a^4
\end{align}
Then we do have a valid microscopic $\Z_2^G$ symmetry, squaring to the identity (note that $a^4$ is in the center of ${\mathbb{D}}_{16}$).  However, the price we pay is that this new $\Z_2^G$ action does not commute with the ${\mathbb{D}}_{16}$ gauge theory Hamiltonian, i.e. the ${\mathbb{D}}_{16}$ analogue of the Hamiltonian written down in equation \ref{ham_2d}.  Indeed, it fails to commute with the ${\mathbb{D}}_{16}$ analogues of the plaquette terms $B''_{v,g}$.  However, if we define a slightly modified plaquette term by:

\begin{align}
B'_{v,g} = \,\, &1 \,\,\,\, \text{ if } h_v^{1} h_v^{\sigma} = 1,a^4 \\
& 0 \,\,\,\, \text{ otherwise},
\end{align}
and analogously for $B'_{l,g}, B'_{p,g}$, and define the vertex terms $A'_{v,g}$ just as we did those for ${\mathbb{D}}_{8}$ in the previous subsection, then the Hamiltonian:

\begin{align} \label{ham_tilde_2d}
H_{\mathbb{D}_{16}}=-\sum_{v,g} A'_{v,g} - \sum_{v,g} B'_{v,g} - \sum_{l,g} B'_{l,g} - \sum_{p,g} B'_{p,g}
\end{align}
is $\Z_2^G$ symmetric.  This Hamiltonian has an extensive ground state degeneracy.  Indeed, the plaquette terms constrain the fluxes through the various plaquettes only up to the $\Z_2$ center of ${\mathbb{D}}_8$, so the ground states correspond to all possible $\Z_2$ flux configurations.  In the next section, we will see how to remove this extensive degeneracy by coupling this model to the surface of a non-trivial 3d SET.  We now give the explicit construction of this 3d SET.

\section{Bulk 3d SET with symmetry fractionalization along gauge flux lines} \label{sec:bulk}

In this section, we write down gapped 3d Hamiltonian with the topological order of a $\Z_2$ gauge theory - i.e. that of the $3d$ toric code, which will realize the anomalous ${\mathbb{D}}_{16}$ theory as a surface state.  This 3d model has pointlike excitations - the $\Z_2$ gauge charges - and looplike excitations - the $\Z_2$ gauge fluxes, or visons.  The model will also have an additional global unitary onsite symmetry $G = \Z_2^G =\{1,\sigma \}$, and is designed to exhibit a special type of symmetry fractionalization which involves the gauge flux loops rather than the gauge charges.  Indeed, as opposed to the more well known scenario of pointlike excitations carrying fractional or projective symmetry quantum numbers, in our model the flux loops transform under $G$ in the same way as edges of a 2d $\Z_2^G$ symmetry protected phase (SPT).  Our model is related to several other 3d models, and we will discuss these connections later.  First, we write down our Hamiltonian.

\subsection{Bulk Hilbert space and Hamiltonian}

We will construct a generalized spin model in 3d, whose degrees of freedom are all spin $1/2$'s.  We will have two such spin $1/2$'s located on each face, edge, and vertex of the 3d cubic lattice.  In order to conform to standard discrete gauge theory terminology, we will refer to faces and edges as `plaquettes' and `links' respectively.  Also, for brevity we will denote the Pauli operators $\sigma^{x}, \sigma^{z}$ simply as $X$ and $Z$.  The two spin $1/2$'s on each plaquette, link, and vertex will be distinguished with a superscript $g=1,\sigma$.  Thus we will be working with the collection of operators

\begin{equation}
\{ X^g_p, Z^g_p,  X^g_l, Z^g_l, X^g_v, Z^g_v \}
\end{equation}
where $g=1,\sigma$ and $p,l,v$ refer to plaquettes, links, and vertices respectively.  The global action of the $\Z_2^G$ symmetry $G$ simply exchanges the two spins:
\begin{equation}
U_{\sigma} X^{g}_p U_{\sigma}^{-1}=X^{g\sigma}_p
\end{equation}
and similarly for the other operators.  The Hamiltonian is:

\begin{equation} \label{ham1}
H=-\sum_{g=1}^{\sigma} \left( A_0^g +A_1^g + B_0+B_1^g+B_2^g+B_3^g   \right).
\end{equation}
Here
\begin{align}
A_0^g &= \sum_{\rm{vertices}\,v} A_{0,v}^g \\
A_1^g &= \sum_{\rm{links}\,l} A_{1,l}^g,
\end{align}
with
\begin{align}
A_{0,v}^g &= Z^1_v Z^{\sigma}_v \prod_{l \sim v} Z^g_l \\
A_{1,l}^g &= Z^1_l Z^{\sigma}_l \prod_{p \sim l} Z^g_p,
\end{align}
where $l \sim v$ means that the product is taken over all $6$ links $l$ that end on $v$, and $p \sim l$ means that the product is taken over all $4$ plaquettes that contain $l$ as an edge.  Also,
\begin{align} \label{defb}
B_0 &= \sum_{\rm{vertices}\,v} B_{0,v} \\
B_1^g &= \sum_{\rm{links}\,l} B_{1,l}^g \\
B_2^g &= \sum_{\rm{plaquettes}\,p} B_{2,p}^g \\
B_3^g &= \sum_{\rm{cubes}\,c} B_{3,c}^g,
\end{align}
with
\begin{align} \label{defb}
B_{0,v} &= \theta X^1_v X^{\sigma}_v  \\
B_{1,l}^g &= X^1_l X^{\sigma}_l \prod_{v \sim l} X_v^g \\
B_{2,p}^g &= X^1_p X^{\sigma}_p \prod_{l \sim p} X_l^g \\
B_{3,c}^g &= \prod_{p \sim c} X_p^g
\end{align}
where the notation is as follows: $v \sim l$ refers to the two endpoints of $l$, $l\sim p$ refers to the 4 edges of $p$, and $p \sim c$ refers to the 6 faces of $c$.  Here $\theta$ is a crucial parameter: for $\theta=1$ the model exhibits no symmetry fractionalization, whereas for $\theta=-1$ we will see that it exhibits symmetry fractionalization along gauge flux loops.  We will set $\theta=-1$ in the remainder of this paper.  

We can consider the Hamiltonian in eq. \ref{ham1} for a variety of 3d geometries.  For concreteness, let us take a cubic lattice on a large 3d torus $T^3$, i.e. take periodic boundary conditions on the cubic lattice.  To understand this Hamiltonian, note first that all of the terms appearing in it commute.  Indeed, commuting any of the terms in $A_j^g$ past any of the terms in $B_k^{g'}$ or $B_0$ leads only to signs from passing Pauli $X$ operators past Pauli $Z$ operators, and these always come in pairs.  What is more, it has a ground state which is a simultaneous eigenvalue $+1$ eigenstate of all of the terms in the Hamiltonian.  An example of such a ground state is given, up to normalization, by:

\begin{align}\label{def_psi}
|\Psi\rangle= \prod_{v} Z_v^1 \prod_{g,v}(1+A_{0,v}^g) \,\prod_{g,l}(1+A_{1,l}^g) |\chi\rangle
\end{align}
where $|\chi\rangle$ is the trivial product state which is the eigenvalue $+1$ eigenstate of all of the $X_v^g$, $X_l^g$, $X_p^g$.  While it is not manifest from its definition in eq. \ref{def_psi}, $|\Psi\rangle$ is actually invariant under $G$.  Indeed, using the fact that $|\Psi\rangle$ is a $+1$ eigenstate of the projectors $(1+A_{0,v}^1)/2$, we have

\begin{align}
&U_{\sigma}|\Psi\rangle = \prod_{v} Z_v^1 Z_v^{\sigma} |\Psi\rangle \\
&=\prod_v \left( \prod_{l\sim v} Z_l^1 \right) |\Psi\rangle = |\Psi\rangle
\end{align}
where the last equality follows since each $Z_l^1$ appears in the product exactly twice.

$|\Psi\rangle$ is not the only ground state when we put the system on the 3d torus $T^3$.  Indeed, our model actually exhibits $\Z_2$ topological order in its bulk.  This can be seen by imposing the terms $B_0$, $B_1^g$, $B_2^g$ as constraints (arguing that this does not remove any topological excitations) to reduce the Hamiltonian in eq. \ref{ham1} to the 3d toric code on the lattice dual to that of the original cubic one.  Thus, for example, two well separated $\Z_2$ gauge charges can be created e.g. by $\prod_p Z^1_p$ where product is along plaquettes bisected by the string connecting the two charges, and a $\Z_2$ gauge flux loop can be created by e.g. $\prod_p X^1_p$ with the product over plaquettes in a membrane which bounds the flux loop.

The hallmark of our model is that the gauge flux loops exhibit a special type of symmetry fractionalization, namely that they behave under the symmetry the same way as edges of $\Z_2$ symmetry protected topological phases (SPTs).  To demonstrate this fact, it is useful to first explain the geometric origin of our model, and rederive some of the above properties in this geometric context.
 
\subsection{Geometric motivation}

In this subsection we explain the geometrical origin of the Hamiltonian in eq. \ref{ham1}, and use what we learn to derive some of its properties.  Underlying our construction is a certain 2-complex $\Gamma$, which is just a generalization of the notion of a graph: in addition to one dimensional links, it also contains two dimensional plaquettes.  Our Hamiltonian can then be thought of as describing 2d membranes moving on this 2-complex, and is in fact a higher dimensional generalization of a construction of Hermele in 2d that involves 1d strings moving on a graph \cite{Hermele_string_flux}.

$\Gamma$ is defined as follows.  Its vertices are simply those of two copies of a cubic lattice, which we think of as being separated in a fictitious fourth dimension.  We will refer to these two copies as the $1$ copy and the $\sigma$ copy, so that for every vertex $v$ in the original cubic lattice, we now have two vertices $v^1$ and $v^{\sigma}$, offset only in the fourth dimension.  For every link $l$ of the original cubic lattice we also have two corresponding links $l^1$ and $l^{\sigma}$.  It will actually be useful later to have an orientation on these links, which is dictated by an arbitrarily fixed orientation on the links of the original cubic lattice.  Additionally, for each $v$ we introduce two new links connecting $v^1$ and $v^{\sigma}$: $l_v^1$, oriented from $v^1$ to $v^{\sigma}$, and $l_v^{\sigma}$, oriented from $v^{\sigma}$ to $v^1$.  Note that these are two distinct links, even though they connect the same pair of vertices.  

\begin{figure*}[htbp]
\begin{center}
\includegraphics[width=0.7\textwidth]{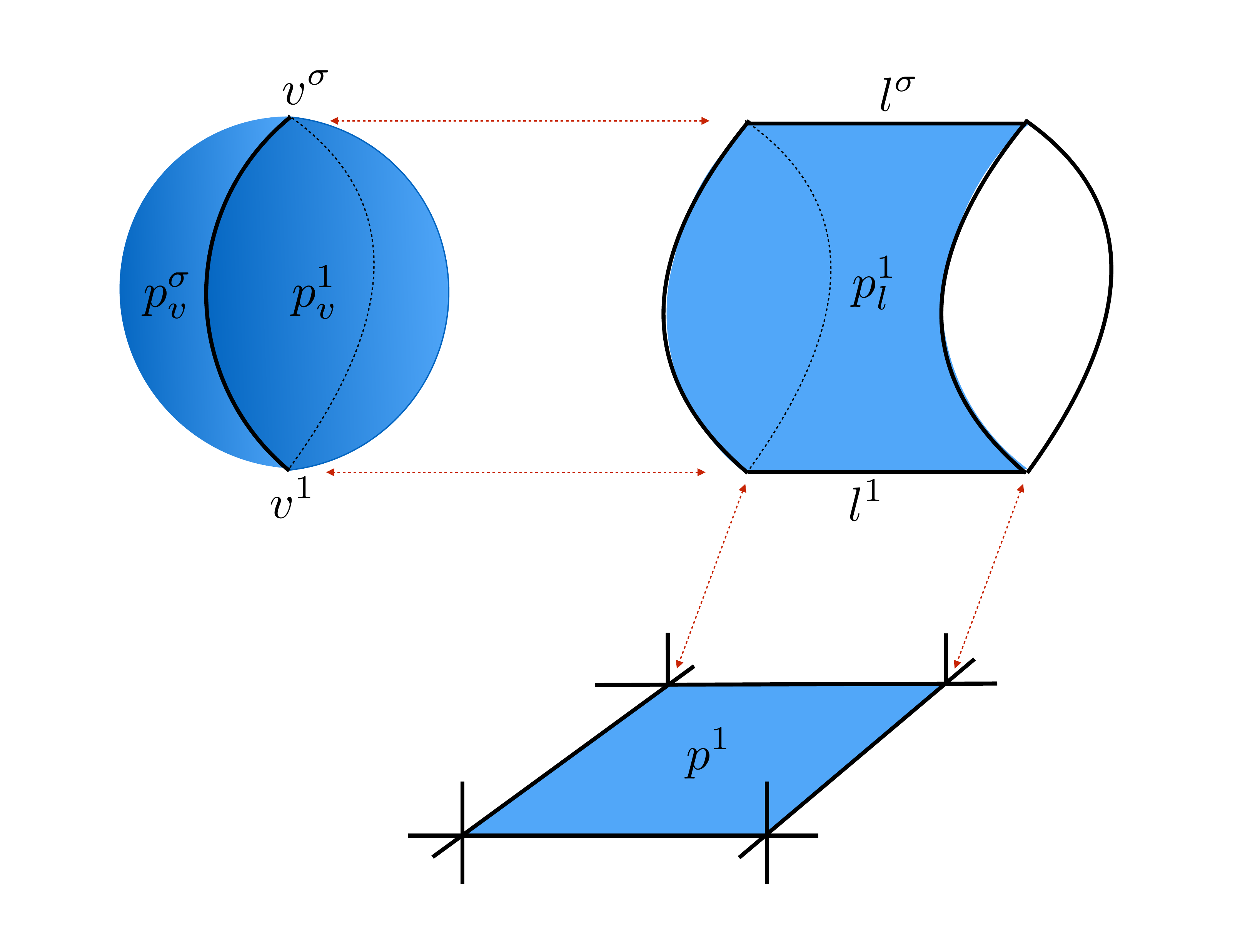}
\caption{The 3 types of plaquettes in our model.  The first type is shown in the lower right, and is simply a square plaquette $p^g$ in one copy ($g=1,\sigma$) of the cubic lattice.  The Pauli operators corresponding to the spin that lives on this plaquette are denoted $X_p^{g}$, $Z_p^{g}$.  The second type is shown in the upper right, and involves a link $l$ of the cubic lattice.  For each such link there are two plaquettes $p_l^1$, $p_l^{\sigma}$ of the second type corresponding to the front and back faces of the cylinder shown at the bottom; we have shown only the front face for clarity.  The Pauli operators corresponding to these are $X_l^1, Z_l^1$ and $X_l^{\sigma}, Z_l^{\sigma}$.  Finally, for each vertex $v$ of the cubic lattice, there exist 2 degenerate 2-sided plaquettes $p_v^1$ and $p_v^\sigma$, which are illustrated as two hemispheres that together form the sphere on the upper left.  These degenerate plaquettes host spins which are acted on by Pauli operators $X_v^1, Z_v^1$ and $X_v^{\sigma}$ and $Z_v^{\sigma}$.
   }
\label{plaq1}
\end{center}
\end{figure*}

As for plaquettes, there are $3$ types (see figure \ref{plaq1}).  First, for each plaquette $p$ of the original lattice, we have two plaquettes $p^1$ and $p^{\sigma}$ in the two copies. Second, for each link $l = \la v_1 v_2 \ra$ of the original lattice, we have two corresponding plaquettes: $p_l^1$, bordered by $l^1$, $l^{\sigma}$, $l_{v_1}^1$, and $l_{v_2}^1$, and $p_l^{\sigma}$, bordered by $l^1$, $l^{\sigma}$, $l_{v_1}^{\sigma}$, and $l_{v_2}^{\sigma}$.  Third, for each vertex $v$, we have two degenerate plaquettes $p_v^1$, $p_v^{\sigma}$, both bordered by $l_v^1$ and $l_v^{\sigma}$.

A particularly appealing feature of $\Gamma$ is that it makes the realization of the symmetry geometrical: the $\Z_2$ symmetry simply acts by switching $g \rightarrow \sigma g$ for all links, vertices, and plaquettes in $\Gamma$.

Now, we can view plaquettes as being occupied or not occupied depending on the eigenvalue of the corresponding Pauli $Z$ operator (with $Z=-1$ being occupied).  Thus a general state in the Hilbert space can be thought of as a superposition of membrane configurations on $\Gamma$ defined by the occupied plaquettes.  The Hamiltonian in eq. \ref{ham1} is simple to understand in this membrane interpretation: the terms $A_0^g$ and $A_1^g$ ($g=1,\sigma$) simply force an even number of occupied plaquettes adjacent to any link of $\Gamma$, reducing us to states supported by closed membrane configurations.  The terms $B_i^g$, on the other hand, force fluctuations of these closed membrane configurations.  Because all the terms in the Hamiltonian commute, we know that the ground state is an eigenstate of each term separately.  However, as we saw above, the model is also unfrustrated, so that the ground state eigenvalue of each such individual term is minimal.  An explicit example of such a ground state was given in eq. \ref{def_psi}; in appendix \ref{unfrust} we give a more geometric construction of this unfrustrated ground state.

\subsubsection{Symmetry fractionalization along gauge flux lines}

The key feature of our model is that it exhibits symmetry fractionalization along the gauge flux lines, in that the gauge flux lines behave like edges of 2d SPTs of $G=\Z_2$.  This can be seen directly from the geometric membrane formulation of our model: the 2d membranes themselves behave like 2d SPTs of $G$.  To see this, we will recast the construction above in a more familiar language, and see that it is related to a particular discretization of a certain non-linear sigma model.

First of all, consider the portion of $\Gamma$ above a particular vertex $v$.  It consists of two vertices, $v^1$ and $v^{\sigma}$, two links connecting these, $l_{v}^1$ and $l_v^{\sigma}$, and two plaquettes $p_v^1$ and $p_v^{\sigma}$ interpolating between these.  To these we can add two 3-cells $c_v^1$ and $c_v^{\sigma}$ interpolating between the plaquettes, corresponding to the $B_{0,v}$ term.  This is nothing but a cellular decoposition of the 3-sphere $S^3$: $v^1$ and $v^{\sigma}$ can be thought of as two poles, whereas $c_v^1$ and $c_v^{\sigma}$ as two (3-dimensional) hemispheres.  The $\Z_2^G$ symmetry then just acts as antipodal reflection: if we think of $S^3$ as composed of unit 4-vectors $\vec{n}$, then $\Z_2$ acts by $\vec{n} \rightarrow -\vec{n}$.

Consider now a single closed membrane $m$; for concreteness say it has the topology of a $2$-sphere $S^2$.  $m$ moves in the space $\Gamma$, but if we think of its 3d position as fixed, then its remaining degrees of freedom are position dependent and take values in the 3-sphere $S^3$ described above.  In other words, the membrane $m$ is described by a sigma model living on $m$.  The important thing about this sigma model is that fluctuating through a 3-dimensional hemisphere $c_v^1$ or $c_v^{\sigma}$ brings a factor of $\theta=-1$, according to eq. \ref{defb}.  Thus, this is just a discretization of a non-linear sigma model (NLSM) with action:

\begin{align}
S&=\int d^2x d\tau \frac{1}{g}(\partial_\mu {\vec{n}})^2 \\ &+\frac{2\pi i}{12\pi^2} \epsilon_{abcd}\epsilon_{\mu\nu\rho} n^a \partial_\mu \phi^b \partial_\nu \phi^c \partial_\rho \phi^d
\end{align}
The $\theta$ term of $2 \pi$ is a reflection of the fact that fluctuating through a 3-dimensional hemisphere brings a factor of $-1$.  Together with the symmetry action $\vec{n} \rightarrow -\vec{n}$, this is just the sigma model description of a $\Z_2$ SPT \cite{Bi2013}.  Thus closed membranes $m$ behave as edges of $2d$ $\Z_2$ SPTs, and therefore gauge fluxes, which are just edges of such membranes, behave as edges of $2d$ $\Z_2$ SPTs.

So far we have just discussed the bulk of our 3d model.  In the next section, we will examine its surface, and couple it to the anomalous 2d theories described in the previous section.

\section{Realizing anomalous anyonic symmetry at the surface of a 3d SET} \label{sec:combined}

The most natural surface termination for the above model is to retain the exact same `$B$' terms, and modify the `$A$' terms for links on the surface in such a way that they still project onto the closed membrane configurations.  For this choice of termination, the $\Z_2$ gauge charges have condensed on the surface: indeed, a single $\Z_2$ gauge charge can be created at the surface by acting with the Pauli $Z$ operator on the spin associated to any plaquette on this surface.  However, 3d $\Z_2$ gauge theory has another natural termination that is dual to this one.  More precisely, in this second sort of termination, vison loops can end on the surface without any confining energy cost coming from the surface \footnote{Because there is a linear confining energy coming from the bulk of the flux loops, a more accurate statement is that there exist U-shaped flux loops which end at two points on the surface, such that the reduced density matrix away from the core of the loops looks like the ground state.  Note that this is not the case for the first surface termination we discussed, which has a surface Higgs condensate.}.  We will see that in order to achieve this second sort of termination in a $G$-symmetric way in our model, we are forced to have topological order on the surface.

To achieve this second sort of surface termination, we first let $S$ denote a surface of the half-infinite cubic lattice; $S$ is then simply a square lattice.  We then define our termination by removing from the Hamiltonian all of the `$A$' terms associated to surface links.  More specifically, we remove the terms in $A_0^g$ corresponding to vertices $v\in S$, and terms in $A_1^g$ corresponding to links $l \in S$.  This allows membranes to end at the surface $S$ with no energy cost; however, it also introduces an extensive ground state degeneracy.  Indeed, any boundary condition on the membranes at $S$, which is just a $\Z_2$ loop configuration on the links of $S$, corresponds to an allowed ground state.  To reduce this degeneracy, we would like to introduce terms that make these loops fluctuate.  A natural choice would be to include the following terms in the Hamiltonian:

\begin{align} \label{extra_term}
- \sum_{g=1,\sigma} \left( \sum_{p \in S} X_p^g + \sum_{l \in S} X_{l}^g + \sum_{v \in S} X_{v}^g \right)
\end{align}
where the sum is over plaquettes $p$ in $S$.  However, even though the inclusion of the term in eq. \ref{extra_term} still results in a commuting Hamiltonian, it is now frustrated.  This is because the terms in $B_0$ corresponding to vertices $v\in S$ want to force $X_{v}^1 X_{v}^{\sigma}=-1$, whereas the terms in eq. \ref{extra_term} prefer $X_{v}^g=1$ for both $g=1,\sigma$.  This frustration leads to an extensive degeneracy of ground states, corresponding to a choice for each $v\in S$ of $g_v$ for which $X_{v}^{g_v} = -1$.  For convenience we will work in the remainder of the paper with the 3d Hamiltonian where we drop the frustrated terms:

\begin{align} \label{ham3d}
H_{3d}=H - \sum_{g=1,\sigma} \left(\sum_{p \in S} X_{p}^g + \sum_{l \in S} X_{l}^g \right)
\end{align}
We will now see that the extensive degeneracy in equation \ref{ham3d} can be cancelled against that of the 2d Hamiltonian in \ref{ham_tilde_2d} realizing the anomalous anyonic symmetry.

\subsection{Coupling surface and bulk}

Our coupled surface-bulk system will consist of the 3d theory defined on the 2-complex $\Gamma$ together with the 2d ${\mathbb{D}}_{16}$ gauge theory degrees of freedom defined on $\Gamma^s$.  Recall that $\Gamma^s$ is the graph defined in the previous section, on which the ${\mathbb{D}}_{16}$ gauge theory was defined.  Now, $\Gamma$ describes a 3d system with a surface, which for definiteness we take to be the $xy$ plane at $z=0$, and $\Gamma^s$ can naturally be thought of as the sub-complex of $\Gamma$, consisting of all the vertices and links at $z=0$.  We then take

\begin{align}
H_{\rm{total}}=H_{3d}+H_{{\mathbb{D}}_{16}} + H_{\rm{interaction}}
\end{align}
where

\begin{align} \label{Hint}
H_{\rm{interaction}}=&-\sum_{v} C_v (X_v^1 - X_v^{\sigma}) - \sum_{l,g} C_{l,g} X_l^g \\ &- \sum_{p,g} C_{p,g} X_p^g -\sum_{v,g} D_v^g Z_v^1 Z_v^{\sigma} \prod_{v' \sim v} Z_{\la v v' \ra}^g \\ &- \sum_{l,g} D_l^g Z_l^1 Z_l^{\sigma} \prod_{p \sim l} Z_p^{g}.
\end{align}
Here $C_v$ is defined as the analogue of the Pauli $Z$ operator for the center $\{1,a^4 \}$ of ${\mathbb{D}}_{16}$ associated with the flux through the plaquette composed of the two edges $l_v^1, l_v^{\sigma}$:
\begin{align}
\rm{eigenvalue}\,\,\rm{of}\,\,C_v = \,\, &1 \,\,\,\, \text{ if } {\tilde{h}}_v^{1} {\tilde{h}}_v^{\sigma} = 1 \\
= &-1 \,\,\,\, \text{ if } {\tilde{h}}_v^{1} {\tilde{h}}_v^{\sigma} = {\tilde{a}}^4 \\
= & 0 \,\,\,\, \text{ otherwise},
\end{align}
Similarly, $C_{l,g}$ and $C_{p,g}$ are analogues of the Pauli $Z$ operators for the center $\{1,a^4 \}$ of ${\mathbb{D}}_{16}$ associated with the flux through the plaquettes $p_l^g$ and $p^g$ respectively.  Also, $D_v^g, D_l^g$ are defined as the analogues of the Pauli $X$ operators for the center of ${\mathbb{D}}_{16}$ associated with the links $l_v^g$ and $l^g$.  They act on the ${\mathbb{D}}_{16}$ labels by:
\begin{align}
D_v^g: & h_v^g \rightarrow a^4 h_v^g\\
D_l^g: & h_l^g \rightarrow a^4 h_l^g.
\end{align}
Essentially, what is happening is that the ${\mathbb{D}}_{16}$ gauge theory Hamiltonian does not touch the $\Z_2$ degrees of freedom corresponding to the center of ${\mathbb{D}}_{16}$, and these can be gapped out against the $\Z_2$ degrees of freedom coming from the bulk.  The anomalous nature of the bulk - i.e. the value of $\theta=-1$ in eq. \ref{defb} - is balanced by the anomalous $\Z_2$ transformation rule in the ${\mathbb{D}}_{16}$ gauge theory in eq. \ref{newGaction}.

Note that the last term in the interaction Hamiltonian in eq. \ref{Hint} can be thought of as condensing the composite object consisting of the $\Z_2 \subset {\mathbb{D}}_{16}$ surface gauge flux and the $\Z_2$ 3d bulk gauge charge.  Hence these two are identified.  At the same time the surface ${\mathbb{D}}_{16}$ charges which transform non-trivially under the $\Z_2$ center of ${\tilde{H}}$ are then bound to endpoints of the 3d bulk vison lines.

\section{Generalizations} \label{sec:gen}

In this paper we have have constructed an example of a discrete gauge theory of the nonabelian group ${\mathbb{D}}_{16}$, together with a permutation action of a group $G=\Z_2$ on its quasiparticles.  We showed that $G$ is anomalous, i.e. cannot be an onsite symmetry in any purely $2d$ realization of this non-abelian gauge theory, but that it can be a symmetry if our non-abelian gauge theory is realized at the center of a symmetry enriched topological (SET) phase.  The price to be paid is that certain quasiparticles of the gauge theory are bound to endpoints of flux lines in the 3d SET, and the signature of the anomaly is that the flux lines exhibit symmetry fractionalization, in the sense of behaving like the edge of a 2d $G$ SPT.

Let us discuss this phenomenon in the context of a general discrete non-abelian gauge theory with gauge group $H$ and center $Z$.  Let us take a general $G$ to act on $H$ by automorphisms, modulo inner automorphisms.  It is easy to generalize both the 2d surface and 3d bulk lattice models to this situation; in particular, the $3d$ bulk becomes a $Z$ gauge theory.  Again, the signature of the anomaly is the symmetry fractionalization of $G$ in along the $Z$ gauge flux lines.  In a $Z$ gauge theory this is generally given by a cohomology class in $H^3(G,Z)$, and this cohomology class is the same as the obstruction class derived in the 2d theory.  Once again, in this realization the irreducible representations of $H$ corresponding - i.e. $H$ charges - in which the center $Z$ acts non-trivially are bound to the ends of $Z$ gauge flux lines.  The same goes for dyons; note that for these, we have to consider the irreducible representations of an appropriate normalizer, but this normalizer always contains the center $Z$.

We have constructed one possible surface termination for the 3d bulk SET in question.  It is natural to ask if there exist any other symmetric ones.  In fact, one simple symmetric termination is given by condensing the $Z$ charges on the surface.  This leads to a surface Higgs condensate, and confines the endpoints of $Z$ gauge flux lines - i.e. the gauge fluxes now cannot end on the surface, and instead must be closed loops in the 3d bulk.  If, however, one insists on not condensing the $Z$ charges at the surface - for example, they may carry fractional charges of some other symmetry group $G'$ which one may not want to break - then one is forced into having topological order at the surface.

However, another puzzle arises from this discussion: the 3d bulk $\Z_2$ SET we have constructed can be obtained in another way by starting with an appropriate 3d bulk SPT of the symmetry group $\Z_2 \times \Z_2^G$ and gauging the $\Z_2$'s (to see this, one can just gauge the remaining $\Z_2^G$ and compare the three loop braiding statistics\cite{WangLevin}).  However, we know that this 3d SPT supports a semion surface state (that is, one with the topological order of a chiral spin liquid), with the semion transforming in a certain projective way under the $\Z_2 \times \Z_2^G$ \cite{ProjS}.  Gauging the $\Z_2$ in this case results in another surface state for our SET, equivalent to the even sub-sector of $U(1)_8$, and the ungauged $\Z_2^G$ acts on this surface theory by reversing the $U(1)_8$ charges (i.e., the fusion algebra of $U(1)_8$ is $\Z_8$, and $G$ acts by $m \rightarrow -m$ for $m\in \Z_8$).

One might be tempted to conclude that the symmetry $m \rightarrow -m$ in $U(1)_8$ suffers from the $H^3$ obstruction anomaly, and cannot be realized in 2d.  However, this is not correct.  Indeed, a simple 2d realization of this symmetry comes from taking the ordinary chiral spin liquid, restricting the symmetry group to $\Z_2 \times \Z_2^G$, and gauging the $\Z_2$.  Upon this gauging, both $\Z_2$ gauge charges and fluxes are introduced, and, together with the semion, these are easily seen to have the structure of $U(1)_8$; furthermore, the other, un-gauged $\Z_2$ then acts by $m \rightarrow -m$.  Thus, even without condensing the 3d bulk $\Z_2$ gauge charges at the surface, our 3d SET does in fact admit a non-anomalous surface.

To resolve this apparent paradox of having both anomalous and non-anomalous surface terminations of the same 3d SET, it is important to first understand precisely what the bulk to boundary anomaly correspondence must be.  The invariant that characterizes the bulk SET is in $H^3(G,Z)$.  However, the surface $H^3$ obstruction is valued in $H^3(G,{\cal{A}})$, where ${\cal{A}}$ is the subgroup of abelian anyons on the surface \cite{ENO, Maissam2014} \footnote{This is only true when the 2d anyon theory is modular}.  These are not the same group; rather there is only a linear map:

\begin{equation} \label{eq:map}
H^3(G,Z) \rightarrow H^3(G,{\cal{A}})
\end{equation}
induced by the embedding $Z \rightarrow {\cal{A}}$, which is just the identification of bulk $Z$ gauge charges with the appropriate surface anyons.

\begin{figure*}[htbp]
\begin{center}
\includegraphics[width=0.6\textwidth]{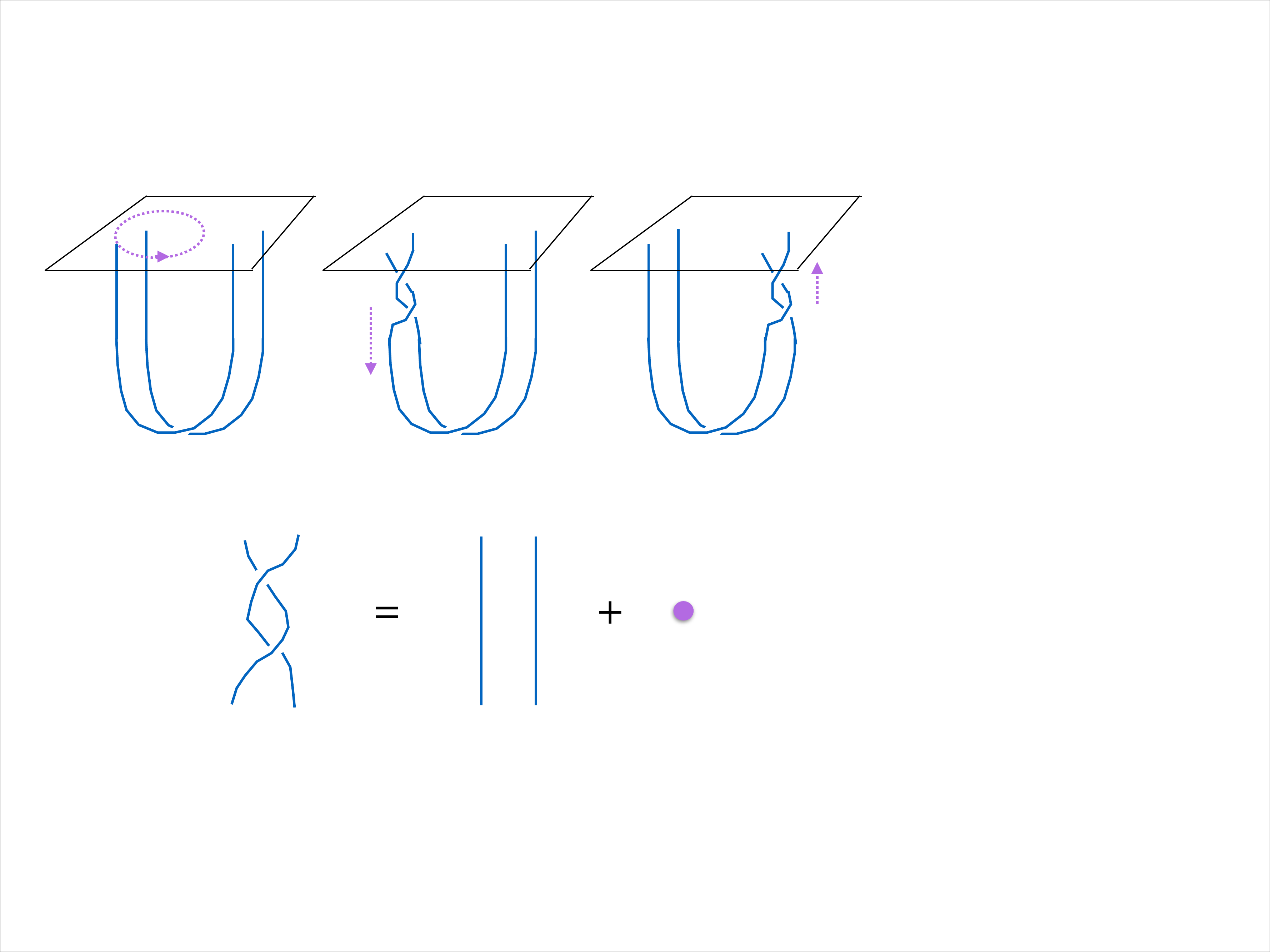}
\caption{The blue lines denote $\Z_2^G$ symmetry fluxes in 3 dimensions; their endpoints are located at the 2d surface.  The purple dot denotes a $\Z_2$ gauge charge, and in our 3d SET, a kink involving two such symmetry fluxes can be undone at the expense of creating such a gauge charge.  Thus the braiding of two symmetry fluxes at the surface transports such a gauge charge from one pair of endpoints to the other.  In a purely 2d theory, this would be an indication of an anomaly.}
\label{fig2}
\end{center}
\end{figure*}

We conjecture that in general the bulk and surface must be such that the fractionalization class of the bulk maps to the surface obstruction class under the map in equation \ref{eq:map}.  In all of the discrete gauge theory examples above, this map is one to one, so that the surface of this form is always non-trivial for a non-trivial SET.  However, in general this is not always so, and, in particular, for our even sub-theory of $U(1)_8$ example above, ${\cal{A}}=\Z_8$, with $Z$ embedded as the subgroup $\{0,4\}$.  For $G=\Z_2^G$, the induced map on cohomology given in equation \ref{eq:map} is zero, so the bulk fractionalization class and surface obstruction again match up, as expected.  Physically, we may interpret the additional anyons in ${\cal{A}}$ as `screening' the obstruction and removing the anomaly.

Let us also mention a physical picture for understanding how the fractionalized bulk `cures' the surface anomaly.  First, consider the 3 loop braiding process of two symmetry flux loops in the presence of a base gauge flux loop in the 3d bulk of our original 3d model (the one with $G=\Z_2^G, Z=\Z_2$) - see figure \ref{fig2}.  This braiding process incurs a Berry's phase of $-1$, which can be understood as follows [cite Xiaoliang]: perform the 3 loop braiding by braiding only a small portion of one symmetry flux loop all the way around the other, and then extend this portion over the whole symmetry flux loop.  This process just propagates a kink (one symmetry flux wrapped around the other) around the whole loop, and the Berry's phase of $-1$ that this process incurs in the presence of a base gauge flux loop means that each kink binds a $Z$ gauge charge.

Now consider the surface, and assume that the $Z$ charges are not condensed, so that we can take a configuration with a $U$ shaped $Z$ gauge flux that ends, in two far separated points, on the surface.  In fact, we can imagine the vacuum to be the trivial 3d SET if we wish, and can complete the $Z$ gauge flux to a full loop.  Carrying out the above process again for two such $U$ shaped $Z$ gauge fluxes next to each other (now there is no base loop, just the two symmetry fluxes), we see that as the kinks propagate, they transfer charge from one endpoint of the $U$ to the other.  Thus, $Z$ gauge charge can be created by simply braiding symmetry fluxes.  This matches up to the surface $H^3$ obstruction because the latter can be thought of as a $Z$ gauge charge ambiguity in the associativity of three $G$ symmetry fluxes, which maps to a $Z$ gauge charge ambiguity in braiding of two such symmetry fluxes, in the same way that a co-cycle in $H^3(G,U(1))$ which defines a 2d $\Z_2^G$ SPT can be understood in terms of the $U(1)$ Berry phases produced by braiding symmetry fluxes \cite{WangLevin}.

Now, in the case of the $m \rightarrow -m$ permutation symmetry of $U(1)_8$, there is clearly no topological charge being created during the full braiding of two $\Z_2^G$ symmetry fluxes, since this symmetry can be realized in an onsite manner purely in 2d.  However, when this theory is realized at the surface of our 3d SET, there is topological charge being created under such a braiding process: namely the $m=4$ quasiparticle is created.  To reconcile these two facts, we note that the creation of such an $m=4$ particle under such a braiding process can be screened by redefining topological superselection sectors associated to the symmetry fluxes.  Specifically, we can bind an $m=1$ anyon to each such symmetry flux.  Then a full braid of two symmetry fluxes turns these into a pair of $m=-1$ anyons, which is a difference of $m=4$.  Thus the existence of $\frac{1}{4}$ fractions of the bulk gauge charge at the surface allows one to screen the anomaly.  More precisely, what we are actually doing is redefining the fusion rules of the symmetry fluxes: we want to deform the fusion rule of two such symmetry fluxes by $m=-2$.  This deformation is precisely the ${\cal{A}}$-valued $2$-cocycle which trivializes the 3-cocycle $z$ that defines the 3d bulk fractionalization.

\section{Discussion} \label{sec:conc}

In the previous section we presented a general conjecture about how the surface $H^3(G,{\cal{A}})$ obstruction class must match up to the bulk $H^3(G,Z)$ fractionalization class, and illustrated it in the case of non-abelian surface gauge theories.  One shortcoming, however, is that this conjecture is applicable only when the 3-cocycle representing the obstruction class can be taken to be valued in a bosonic subgroup $Z$ of the abelian anyons.  One possibility is that this is always the case, for any surface theory with an $H^3(G,{\cal{A}})$ obstruction.  In this case, $Z$ is always of the form $\rm{Rep}\,Z^*$, and it is known that any such anyon theory comes from a smaller anyon theory with $Z^*$ symmetry through gauging \cite{DGNO}.  Then the story presumably reduces to the $H^4(G\times Z^*,U(1))$ obstruction story for 3d $G\times Z^*$ SPTs.  However, there is no obvious reason to suspect that the obstruction is always valued in such a bosonic subgroup, so the general case is still open.  One partial generalization comes from noting that in a general bosonic 3d bulk SET, the emergent pointlike quasiparticles could also be fermions.  So really we only need the surface obstruction to be valued in a subgroup of the abelian anyons where any two particles are mutual bosons.  Again, it is not known whether or not this is true in general.

The possibility of fermionic quasiparticles in the bulk brings up the exciting possibility of extending our work to fermionic SETs and surface topological order; the conjecture given in equation \ref{eq:map} generalizes naturally to this setting.  In particular, we can take $Z$ to be the $\Z_2$ group whose single non-trivial particle is the electron.  Then the 3d bulk SET can be thought of as a fermionic SPT, and our story turns into that of surface terminations of fermionic SPTs.  One class of such terminations was constructed in reference \onlinecite{Fidkowski2013} for the topological superconductor in Cartan class DIII, and in particular the integral subtheory of $SU(2)_6$ was proposed as a surface termination of such a topological superconductor with an odd value of the index.  Even though the relevant symmetry here, namely time reversal, is anti-unitary, one may imagine that our story extends to this case as well.  Supporting this idea is the fact that the flux lines in the 3d bulk of such a topological superconductor support protected gapless modes (in the mean field free fermion realization of the bulk), and these modes behave like the edges of a fermionic 2d SPT, namely a $p+ip \uparrow / p-ip \downarrow$ superconductor.

It would be interesting to relate our construction to recent work of Kapustin \cite{Kapustin_gen_sym}; see also reference \onlinecite{CurtRyan}.  In particular, these works discuss extending the notion of symmetry from an onsite tensor product action (`0 symmetry') to higher form symmetries.  A Walker-Wang construction of a 3d SPT based on the obstructed 2d modular theory will realize the $G$ symmetry in this generalized higher form fashion, and it would be interesting to see if there are any connections between this and the construction given in this paper. 

\section{Acknowledgements}

We would like to acknowledge useful discussions with Barry Bradlyn, Xie Chen, Pavel Etingof, Anton Kapustin, and Curt von Keyserlingk.  AV was supported by the Simons Investigator program.

\appendix

\section{Obstruction to the existence of a group extension} \label{group_obstruction}

This appendix is based on reference \onlinecite{H3math}.  Let $G$ and $H$ be groups, with $H$ non-abelian and $Z\subset H$ the center of $H$.  Recall the center $Z$ consists of all group elements of $H$ which have the property that they commute with all of $H$.  First, suppose these fit into a short exact sequence:

\begin{equation}
1\rightarrow H \rightarrow {\tilde{G}} \rightarrow G \rightarrow 1
\end{equation}
This just means that $H$ is a normal subgroup of ${\tilde{G}}$ (i.e. one that is mapped to itself when conjugated by any element of ${\tilde{G}}$), and the quotient ${\tilde{G}} / H$ is isomorphic to $G$; the second and third arrows above represent this inclusion of $H$ in ${\tilde{G}}$ and the quotient map respectively.  The existence of this short exact sequence is just the mathematical formalization of the existence of a larger gauge group ${\tilde{G}}$ obtained by gauging the symmetry $G$ in the $H$ gauge theory discussed in the main text.

Now, for every $g\in G$, pick a lift $x_g \in {\tilde{G}}$ - this is just something which maps to $g$ under the quotient map ${\tilde{G}} \rightarrow G$.  Since $\beta(g,h) \equiv x_g x_h x_{gh}^{-1}$ maps to the identity in $G$ under the quotient, it means that it must be an element of $H$.  Furthermore, associativity of multiplication of $x_g, x_h, x_k$ in ${\tilde{G}}$ gives us the condition:

\begin{equation} \label{cocyc_beta}
\beta(g,h) \beta(gh,k) = \left( x_g \beta(h,k) \right) \beta(g,hk)
\end{equation}

The lift $x_g$ we chose is ambiguous up to multiplication by an element of $H$.  Another way to look at this is that conjugation by $x_g$ gives us an automorphism of $H$ that is defined only up to an inner automorphism of $H$ (that is, one given by conjugation by something in $H$).  Thus, it gives us a well defined element ${\gamma_g}$ of $\rm{Out}(H)$, the group of all automorphisms of $H$ modulo the inner ones; also, it is clear that given an element of $\rm{Out}(H)$, we can uniquely back out $x_g$ up to multiplication by an element of $H$.  Furthermore, from the above we have that ${\gamma_f} {\gamma_g} = {\gamma_{fg}}$.  Thus, from a short exact sequence of the above form, we can extract the data of a group map $\gamma: G \rightarrow \rm{Out} (H)$.

Now suppose we reverse the question: given a group map $\gamma: G \rightarrow \rm{Out} (H)$, can we find an extension ${\tilde{G}}$ and a short exact sequence that induces it, in the manner described above?  To answer this, we first pick, for each $g$, an automorphism ${\tilde{\gamma}}_g$ of $H$ in the equivalence class of $\gamma_g$.  Then the automorphism ${\tilde{\gamma}}_g {\tilde{\gamma}}_h \left({\tilde{\gamma}}_{gh} \right)^{-1}$ must be given by conjugation by some element that we will call $\beta'(g,h)$.  Now, in order for $\gamma$ to be induced by an extension ${\tilde{G}}$ and short exact sequence of the above form, we must have $\beta'$ satisfy equation \ref{cocyc_beta}.  However, sometimes this does not happen.  Indeed, even though the composition of automorphisms ${\tilde{\gamma}}_g$ is associative, the co-cycle equation that follows for $\beta'$ holds only up to an element $z(g,h,k)$ of the center $Z$ of $H$:

\begin{equation} \label{cocyc_beta_prime}
\beta'(g,h) \beta'(gh,k) = z(g,h,k) {\tilde{\gamma}}_g(\beta'(h,k)) \beta'(g,hk)
\end{equation}

Here $z(g,h,k) \in Z$.  It can be verified that $z$ is a 3-cocycle, i.e. it satisfies equation \ref{cocyc_z}, and that different choices of ${\tilde{\gamma}}$ and $\beta'$ yield a 3-cocycle $z'$ which is related to $z$ via equation \ref{gauge_z}.  Thus the cohomology class $[z] \in H^3(G,Z)$ is uniquely defined, and, when non-trivial, signals the obstruction to finding a group extension ${\tilde{G}}$ with the above properties.

This discussion simplifies when $G=\Z_2 = {1,\sigma}$.  Indeed, in this case we only have to check whether $z(\sigma,\sigma,\sigma)$ is non-trivial or not.  Furthermore, the only non-trivial $\beta$ is $\beta(\sigma,\sigma)$, which is equal to the element $\alpha$ of $H$ by which one conjugates to obtain the action of $\rho_g^2$, and equation \ref{cocyc_beta_prime} reduces to:

\begin{equation}
z(\sigma,\sigma,\sigma)= \alpha \left( {\tilde{\gamma}}_g (\alpha) \right)^{-1}
\end{equation}
In the case of $H={\mathbb{D}}_8$, we have $\alpha=a^{-1}$, and ${\tilde{\gamma}}_{\sigma}(\alpha)=\alpha$, so that $z(\sigma,\sigma,\sigma)=1$, which is interpreted as `0' in the additive notation we are using for $Z$.  Hence, in this case the obstruction vanishes.  However, in the case of $H={\mathbb{D}}_{16}$ we have $\alpha={\tilde{a}}^{-3}$, and ${\tilde{\gamma}}(\sigma) (\alpha)={\tilde{a}}$, so that $z(\sigma,\sigma,\sigma)={\tilde{a}}^4$ and hence the obstruction is non-trivial.

\section{Unfrustrated ground state} \label{unfrust}

Consider a putative ground state as a superposition of membrane configurations, where each membrane enters with amplitude $s(m)/{\sqrt N}$, where $N$ is the number of closed membrane configurations and the sign $s(m)=\pm 1$ associated to a membrane configuration $m$ is determined by the requirement that all of the terms in $B$ (see eq. \ref{defb}) have eigenvalue $+1$ when acting on this ground state.  The unfrustrated nature of our model is equivalent to the existence of a choice of sign $s(m)=\pm 1$ satisfying this requirement.  To see that such an $s(m)$ exists, we now explicitly define it.  

For this part of the argument, we want to consider a 3d system without boundary, so we take an infinite 3-dimensional space and associated cubic lattice\footnote{A more formal argument would require examining a finite system without boundary.}.  Also, it will be useful to also consider all of the $3$-cells through which membranes of $\Gamma$ fluctuate, i.e. those implicitly defined in eq. \ref{defb}.  Including these $3$-cells results in an augmentation of $\Gamma$ to a larger 3-complex $\Gamma'$.  In fact, we will want a slight variation on this construction, where we add {\emph{two}} 3-cells corresponding to the term $B_0$ in eq. \ref{defb}.  One way to understand $\Gamma'$ is as the 3-skeleton of the space $S^3 \times {\mathbb{R}}^3$.  Let us explain this construction carefully.  

First, the $S^3$ is to be interpreted as the portion of $\Gamma'$ over a single vertex $v$; it consists of two vertices $v^1$ and $v^{\sigma}$, two links $l^1$ and $l^{\sigma}$, two $2$-cells $p_v^1$ and $p_v^{\sigma}$, and two $3$-cells $c_v^1$ and $c_v^{\sigma}$ forming hemispheres of the $S^3$.\footnote{This cellular decomposition of $S^3$ allows us to naturally define a free action of $\Z_2$ (by $g \leftrightarrow \sigma g$ on $v^g, l_v^g, p_v^g, c_v^g$), and is the one arising from the standard construction of the classifying space $B\Z_2$ as the quotient $E \Z_2 / \Z_2$}  The ${\mathbb{R}}^3$ also carries a cellular structure, namely the one given by the vertices, links, plaquettes, and cubes of the cubic lattice.  $\Gamma'$ is then defined by taking the product complex $S^3 \times {\mathbb{R}}^3$, and retaining only the $3$-cells within it.  These $3$-cells come in $4$ types, namely as products of $j$-cells in $S^3$ and $(3-j)$-cells in ${\mathbb{R}}^3$, for $j=0,1,2,3$.  

Two purely topological facts about $\Gamma'$ are crucial to our argument.  The first is that any closed 2d membrane $m$ is the boundary of a union $R_m$ of 3-cells in $\Gamma'$, and is a consequence of the fact that the second homology group of $\Gamma'$ with $\Z_2$ coefficients vanishes.  The second is that any union of 3-cells $R'$ whose boundary is trivial must contain an even number of 3-cells of the form $c_v^g$ (i.e. those of the form $3$-cell in $S^3$ times $0$-cell in ${\mathbb{R}}^3$).  To see this, consider the inclusion of $R'$ in $\Gamma'$ followed by the projection $S^3 \times {\mathbb{R}}^3 \rightarrow S^3$: being a continuous map from a closed $3$-manifold to $S^3$, it must wrap $S^3$ an integral number of times, and since $S^3$ is composed of precisely two $3$-cells of the form $c_v^g$, $R'$ itself must include an even number of such 3-cells.

Given these two facts, we can now unambiguously define $s(m)= -1$ whenever the number of 3-cells of the form $c_v^g$ in a bounding region $R_m$ of $m$ is odd and $s(m)=+1$ otherwise.  This definition of $s(m)$ is independent of the choice of $R_m$ using this the second fact above.  Then the superposition of closed membrane configurations with amplitudes $s(m)/{\sqrt N}$ is manifestly a ground state of all of the individual terms in eq. \ref{defb} - the important fact is that it is precisely the 3-cells of the form $c_v^g$ which enter with a minus sign when $\theta = -1$ in eq. \ref{defb}.  Furthermore, since the $s(m)$ are uniquely determined up to overall sign by the condition that all the terms in eq. \ref{defb} be satisfied, and because all the terms in the Hamiltonian commute, we see that the model is gapped with a unique ground state.

\bibliographystyle{unsrt}
\bibliography{H3refs}

\end{document}